\documentclass[prb,authorgroup,twocolumn,showpacs,preprintnumbers,amsmath,amssymb]{revtex4}

\usepackage{graphicx}
\usepackage{dcolumn}
\usepackage{bm}
     \usepackage{mathrsfs}

\newcommand{\dd}{\textrm{d}}

\newcommand{\ee}{\textrm{e}}

\DeclareBoldMathCommand{\bnabla}{\nabla}
\DeclareBoldMathCommand{\bcdot} {\cdot}
\DeclareBoldMathCommand{\btimes}{\times}

\newcommand{\ud}[1]{\textrm{d}#1\,}

\begin{document}

\preprint{APS/123-QED}

\title{Determination of Trion and Exciton Lineshapes in Modulation-Doped Quantum Wells}

\author{Jean~Berney}
 \affiliation{Institut de Photonique et Electronique Quantiques,
 Ecole Polytechnique F\'{e}d\'{e}rale de Lausanne (EPFL) CH1015
 Lausanne, Switzerland}
 \author{Lucas~Schifferle}
 \affiliation{Institut de Photonique et Electronique Quantiques,
 Ecole Polytechnique F\'{e}d\'{e}rale de Lausanne (EPFL) CH1015
 Lausanne, Switzerland}
\author{Marcia T.~Portella-Oberli}
 \affiliation{Institut de Photonique et Electronique Quantiques,
 Ecole Polytechnique F\'{e}d\'{e}rale de Lausanne (EPFL) CH1015
 Lausanne, Switzerland}
\author{Beno{\^i}t~Deveaud}
 \affiliation{Institut de Photonique et Electronique Quantiques,
 Ecole Polytechnique F\'{e}d\'{e}rale de Lausanne (EPFL) CH1015
 Lausanne, Switzerland}
\date{\today}

\begin{abstract}
We investigate the effect of a two dimensional electron gas on the linear optical properties of CdTe quantum wells. We evidence experimentally the high energy tail of the exciton and charged exciton resonances which depends on electron concentration. Based on that, we show that the scattering of electrons with excitons and charged excitons is needed to be included in the matrix transfer calculations to describe the reflectivity spectra. We demonstrate by time-resolved reflectivity experiments the importance of electron distribution in the resonance lineshapes.
\end{abstract}


\maketitle

\section{Introduction}

Past experimental studies have pictured the linear optical properties of quantum wells in the presence of a moderate background electron density. They showed that the exciton resonance is strongly affected by the electron population. First of all electrons screen the excitons and reduce their oscillator strength \cite{chemla:1985}. Then, they fill up the conduction band and contribute to shift the exciton resonance towards higher energies \cite{huard:2000,kossacki:1999,yusa:2000}. Finally they collide with electrons and broaden their absorption line \cite{huard:2000}. Experimental comparison between exciton-electron and exciton-exciton scattering mechanism in bulk GaAs \cite{schultheis:1986}, and in GaAs quantum wells \cite{honold:1989,capozzi:1993}, showed that the exciton-electron scattering efficiency is one order of magnitude larger than exciton-exciton process. Both scattering processes are enhanced for the two-dimensional excitons as compared to bulk excitons.
 
Linear optical spectrum of modulation-doped quantum wells also features a charged exciton (trion) resonance below the exciton resonance \cite{kheng:1993}. Theoretical calculations of the binding energy of charged excitons were performed using variational \cite{stebe:1989,thilagam:1997,stebe:1997} and full solution of the three-particle Schr{\"o}dinger equation \cite{riva:2000, esser:2001}. They showed that only the singlet state is bound in the absence of magnetic field. Considering the lineshape of the trion resonance, it was argued that the momentum of the electron initially present in semiconductor and used in the absorption process of a trion yields a low-energy tail on the trion resonance \cite{suris:2001, esser:2001}. A simple derivation of the trion-electron scattering was also performed in GaAs quantum wells \cite{ramon:2003}, but the overlap of the exciton and trion resonances did not allow clearing identifying if such simple model leads to correct predictions.

In this paper, we investigate the effect of a two-dimensional electron gas on the optical properties of CdTe quantum wells. Due to the large energy separation between exciton and trion resonances, we are able to evidence experimentally the existence of a high-energy tail of both exciton and trion resonances. An electron-exciton and electron-trion scattering model, taking into account both elastic and inelastic scattering processes, show quantitative agreement with our experimental linear spectra. These findings, even being yet far, can go in the direction of the modern theory of excitons within an electron gas, that argue that a trion is intrinsically a many-body object, made of a hole interacting with all electrons in the system \cite{combescot:2005}. We use also time- and spectrally- resolved pump-probe experiments to investigate the effect of scattering of electrons with trion and excitons on the nonlinear reflectivity spectra.

The paper is structured in the following way: We give the sample characteristics in Sec. \ref{sec:sample}. After presenting the measured reflectivity spectra of the sample in Sec. \ref{sec:experiment}, we show, in Sec. \ref{sec:calculation}, that a calculation without exciton- and trion-electron scatterings cannot possibly describe our system. Sec. \ref{sec:scattering} summarizes the exciton-electron and trion-electron calculation results and we demonstrate that linear optical spectra of modulation-doped quantum wells are quantitatively described by a matrix transfer calculation including both exciton- and trion-electron scatterings. In Sec. \ref{sec:tr-experiment} using time-resolved pump-probe experiment we illustrate the effect of the electron population on the exciton and trion lineshapes.  We end with the conclusion in Sec. \ref{sec:conclusion}.

\section{Sample Structure}\label{sec:sample}

We worked on a one-side modulation doped CdTe/Cd$_{0.27}$Mg$_{0.73}$Te heterostructure \cite{wojtowicz:1998}, containing a single CdTe quantum well of 8 nm between 50 nm thickness barriers. This structure was grown on top of a 4.5 mm of CdTe which is deposited above a GaAs substrate. A remote donor layer of iodine with thikness of 49 \AA was embedded in the cap layer 10 nm apart from the quantum well, yielding a population of about $4 \times 10^{10}$~cm$^{-2}$ excess electrons in the QW. The electrons from the donors can either fall in the quantum well or be trapped by surface states, both processes competing. In this regime, at temperature of 5 K, two strong resonances separated by 3 meV arise below the band gap; they are attributed to heavy-hole excitons (1625.7 meV) and negatively charged excitons (trions, 1622.4 meV). A control of the electron density can be achieved by illuminating the sample with light more energetic than the energy gap of the barrier, i. e. larger than 2.03 eV. In this case, we create electron-hole pairs in the barrier. Holes are attracted by electrons trapped in the surface states, while electrons are repelled towards the quantum well. By increasing the light intensity, we can increase the electron gas density. For instance, the practical electron concentration range from $4.3 \times 10^{10}$~cm$^{-2}$ to $1.2 \times 10^{11}$~cm$^{-2}$ is attained. The thickness of the barriers was chosen in such a way that reflectivity spectra are absorption-like and can be directly interpreted.

\section{ CW Reflectivity Experiments}\label{sec:experiment}

We performed reflectivity measurements of the sample at temperature of 5 K. A white light source was spectrally filtered so that any high energy component ($> 1.7$ eV) likely to modify the density of the electron gas in the quantum well was suppressed. The light was collimated and focused on the sample. The reflection was then spectrally resolved by an imaging monochromator and recorded by a CCD camera. We kept the incident light intensity low and made sure we were working in linear regime.
  The light coming from a blue GaN light emitting diode (LED) was focused on the sample and completely covered the white light spot. By monitoring the intensity of the blue LED we could change the density of electrons in the well. The diffusion time of electrons from the barrier to the quantum well is macroscopic; it usually took 5 s to 15 s for the system to reach equilibrium whenever the LED power was tuned. Reflectivity spectra were recorded for various electron densities.

\begin{figure}[h]
    \includegraphics[scale=0.85]{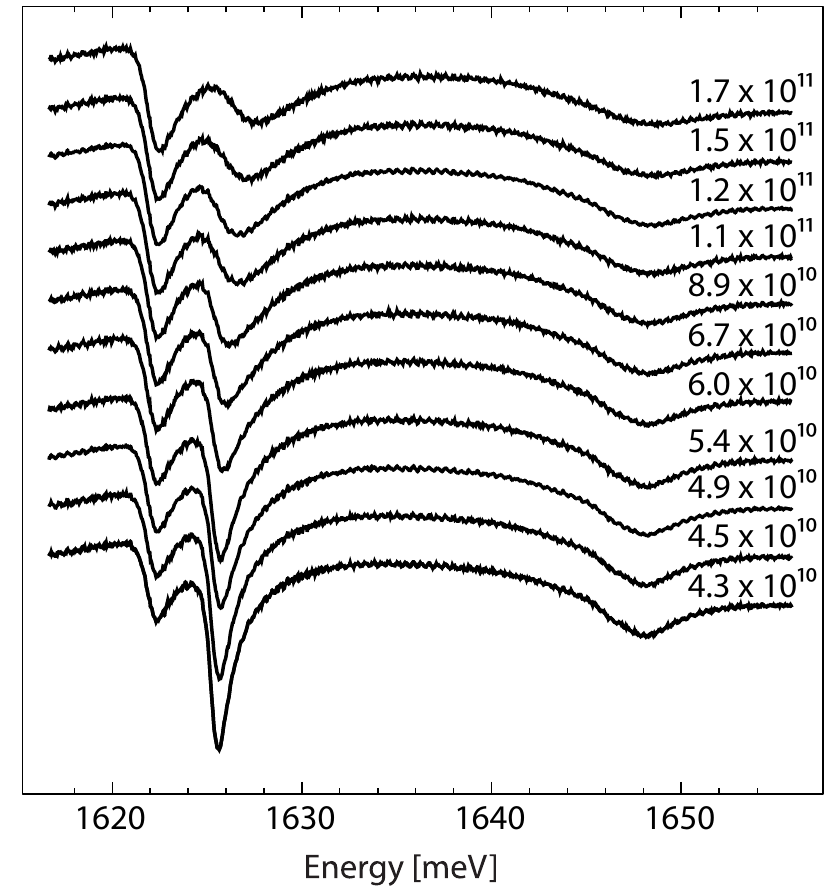}
    \caption{\label{fig:reflectivity}
       CW reflectivity spectra obtained at 5 K for different electron densities ranging from $4.3 \times 10^{10}$~cm$^{-2}$ to $1.7 \times 10^{11}$~cm$^{-2}$.
    }
\end{figure}

Figure 1 shows the reflectivity spectra obtained for different electron densities. For increasing electron densities, we clearly observe a blue shift of the exciton resonance as reported in the literature \cite{huard:2000,kossacki:1999,yusa:2000}. It is due to the filling of the conduction band by electrons. Electron-hole pairs can only be photo-generated above the Fermi level. To a good approximation, the shift is linear in the Fermi energy $E_F$. Thus the exciton-trion energy difference is given by  $E=E_0^T+\alpha E_F$, where $E_0^T$  is the trion binding energy and $\alpha$ is an empirical parameter. For CdTe quantum wells, $\alpha = 1.07$ \cite{huard:2000}. Assuming that $E_0^T= 1.7$ meV, we estimate the electron concentrations in the quantum well for each spectrum in Fig. 1. The densities that we obtained range from $4.3 \times 10^{10}$~cm$^{-2}$ to $1.7 \times 10^{11}$~cm$^{-2}$.

We note that in the measured spectra, also a high energy tail of exciton and trion resonances develops as the electron density increases. We will show that this finding can only be explained by considering electron interactions with excitons and trions. 

\section{Reflectivity Calculation of the Sample}\label{sec:calculation}

As a starting point, we chose to use the linear susceptibility proposed by \cite{esser:2001} in the framework of density matrix theory. It includes both exciton and trion contributions, has a rather simple form and takes into account the nonzero momentum of the excess electron in the quantum well used in the process of trion photogeneration. We emphasize that it neglects any scattering with electrons. It reads

\begin{equation}\label{eq:chi}
\chi(\omega)=\chi^X(\omega)+\chi^T(\omega),
\end{equation}
where $\chi^X(\omega)$ and $\chi^T(\omega)$ are, respectively, the contributions to
the exciton and trion susceptibility. They are given by
\begin{align}\label{eq:chiXT1}
    \chi^X(\omega)=&f_X\frac{|\phi_X(0)|^2}{\omega-\omega_X-i\gamma_X},\\
    \chi^T(\omega)=&f_T\int
    \ud{\bm{q}}n_e(\bm{q})\frac{|M^T(\bm{q})|^2}{\omega-\omega_T+W_{\bm{q}}-i\gamma_T}.\label{eq:chiXT2}
\end{align}
where $\omega_X$ ($\omega_X$) is the exciton resonance energy, $f_X$
($f_T$) a contribution to the exciton (trion) oscillator strength,
and $\gamma_X$ ($\gamma_T$) the exciton (trion) homogenous
broadening mainly due to phonons; $n_e(\bm{q})$ is the Fermi-Dirac
distribution of the electrons and $W_{\bm{q}}=\hbar^2q^2M_X/2m_eM_T$
is a correction to the trion resonance that comprises trion center of mass
energy and electron initial momentum $\bm{q}$, with $m_e$, $M_X$ and
$M_T$ the electron, exciton and trion effective mass, respectively.
For a given initial electron momentum, the strength of the integrant
is weighted by the optical matrix element
\begin{equation}\label{???}
    M^T(\bm{q})=\int\ud{\bm{\rho}_2}\psi^T_{1s}(0,\bm{\rho}_2)e^{i \eta \bm{q} \bm{\rho}_2}
\end{equation}
where $\eta=M_X/M_T$ and $\psi^T_{1s}(\bm{\rho}_1,\bm{\rho}_2)$ is
the trion $1s$ wavefunction, with $\bm{\rho}_1=\bm{r}_{1e}-\bm{r}_h$
the relative position of the first electron to the hole and
$\bm{\rho}_2=\bm{r}_{2e}-\bm{r}_h$ from the second electron to the
hole.

For the trion wavefunction, we relied on a simple variational
function \cite{sergeev:2001} with only two variational parameters $\lambda$ and $\lambda'$. It
is given by
\begin{equation}\label{eq:chandrasekhar}
\psi_T(\rho_1,\rho_2)= \mathcal{N}_T \left(e^{\rho_1/\lambda
-\rho_2/\lambda'} + e^{\rho_1/\lambda' -\rho_2/\lambda} \right),
\end{equation}
Where $\mathcal{N}_T$ is the normalization factor. With this
approximation, the light coupling element $M(\bm{q})$
 becomes
\begin{equation}\label{eq:MT}
    M_T(\bm{q})=4 \mathcal{N}_T\left(\frac{\lambda_T}{\lambda_T'}\frac{1}{(1+(\lambda_T
    \bm{k})^2)^{3/2}}+\lambda_T\leftrightarrow\lambda_T'\right).
\end{equation}

We evaluated, in Appendix C, the error induce by using the Chandrashekar's
wavefunction. In Fig.~\ref{fig:Mtrion}, we compare the result
obtained with variational functions \ref{eq:sergeev} and
\ref{eq:chandrasekhar}. They do not differ much and we conclude that
Eq.~\ref{eq:MT} is a reliable approximation.

Exciton and trion resonance may also be inhomogenously broadened,
due to quantum well interface roughness. Building on
\citet{andreani:1998} work, we supposed that the broadening can be
described by a Gaussian distribution function. Thus, for excitons,
we substituted the dielectric susceptibility (\ref{eq:chiXT1}) by
the convolution function
\begin{align}\label{eq:InhomogenousBroaderingX}
   \widetilde{\chi}^X(\omega)&=\frac{1}{\sqrt{\pi}}\int\ud{\nu}\chi^X(\omega-\nu)\exp\left[-\left(\frac{\nu-\omega_X}{\Gamma_\text{inhom}^X}\right)^2\right]\notag\\
   &=\frac{i\pi f_X |\phi_X(0)|^2}{\Gamma_\text{inhom}^X}w\left(\frac{\omega-\omega_X-i\gamma_X}{\Gamma_\text{inhom}^X}\right),
\end{align}
where $w$ is the complex error function \cite{schreier:1992} and
$\Gamma_\text{inhom}^X$ the exciton inhomogenous broadening
constant. For trions the same kind of equation holds, except the
convolution is performed before integrating over $\bm{q}$:
\begin{align}\label{eq:InhomogenousBroaderingT}
   \widetilde{\chi}^T(\omega)&=
   \frac{i\pi f_T}{\Gamma_\text{inhom}^T}\int
    \ud{\bm{q}}n_e(\bm{q})|M^T(\bm{q})|^2\\
    &w\left(\frac{\omega-\omega_T+W_{\bm{q}}-i\gamma_T}{\Gamma_\text{inhom}^T}\right)
\end{align}

\begin{figure}[htb]
\begin{center}
    \includegraphics[width=0.4\textwidth]{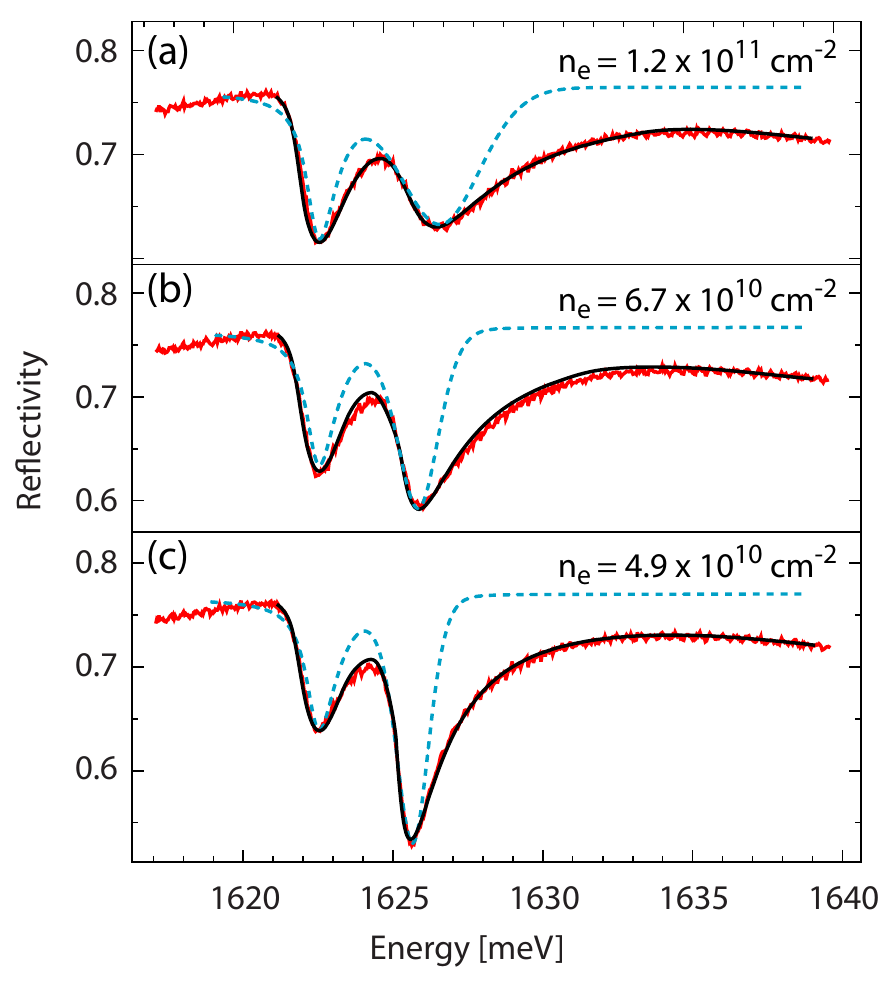}
\end{center}
\caption[cw reflectivity spectra and fits including electron
scattering.]{(red solid) cw reflectivity spectra measured at $5$~K
and for electron densities of $4.9\times10^{10}$, $6.7\times10^{10}$
and $1.2\times10^{11}$~cm$^{-2}$. Best fit obtained with electron
scattering (black solid) and without (blue dashed).}
\label{fig:ReflectivityFit_exchange}
\end{figure}

Using the transfer matrix formalism described in Appendix A and B, we calculated the mode of the field for our sample
structure. The refractive index and absorption of the CdMgTe
barriers \cite{choi:1997,andre:1997}, CdTe quantum well, CdTe buffer
\cite{benhlal:1999,hlidek:2001}, GaAs substrate and iodine dopant
were all implemented. For each resonance, we used the following
fitting parameters: spectral position, oscillator strength,
homogenous linewidth and inhomogeneous linewidth. We fitted three
densities: $4.9\times10^{10}$, $6.7\times10^{10}$,
$1.2\times10^{11}$~cm$^{-2}$. Results are shown in
Fig.~\ref{fig:ReflectivityFit_exchange} and the value of the
corresponding fitting parameters in
Table~\ref{table:FittingParameters}.
\begin{table}[!tb]
\begin{center}
\begin{tabular}{l c c c c c }
\hline
&$n_e$ [cm$^{-2}$]&$E$ [meV]&$\Gamma_{\text{hom}}$ [meV]&$\Gamma_{\text{inhom}}$ [meV]\\ \hline  %
$X$ & $4.9\times10^{10}$ & 1623.9 &  0.25 & 0.1\\
$X$ & $6.7\times10^{10}$ & 1624.1 &  0.15 & 0.1\\
$X$ & $1.2\times10^{11}$ & 1624.4 &  0.10 & 0.1\\
$X^-$ & $4.9\times10^{10}$ & 1621.9 &  0.2 & 0.05\\
$X^-$ & $6.7\times10^{10}$ & 1621.9 &  0.3 & 0.05\\
$X^-$ & $1.2\times10^{11}$ & 1622.1 &  0.3 & 0.10\\
\hline
\end{tabular}
\end{center}
 \caption{Fitting parameters of the reflectivity spectra for three electron densities $n_e$. The homogenous broadening $\Gamma_{\text{hom}}$ is essentially due to phonon interaction and corresponds to $\gamma_X$ for excitons and $\gamma_T$ for trions (see
 Eq.~\ref{eq:chiXT1} and~\ref{eq:chiXT2}). The inhomogenous broadering is defined in Eqs.~\ref{eq:InhomogenousBroaderingX} and~\ref{eq:InhomogenousBroaderingX}.}
 \label{table:FittingParameters}
 \end{table}

From the fits in Fig.~\ref{fig:ReflectivityFit_exchange}, we see
that it is impossible to reproduce both low energy and high energy
tail of exciton and trion resonance because of their strong
asymmetry.
We note that in the measured spectra, the low-energy tail of the trion resonance predicted by \cite{esser:2002} is marginal and can be fairly well reproduced by our fit. Conversely, a high energy tail of exciton and trion resonances develops as the electron density increases. Since we did not take into account the interaction between exciton and trion with electrons we cannot reproduce the correct exciton and trion lineshapes. In the next section, we will use a simple model of exciton- and trion-electron interaction that will fit quite well the high-energy broadening of measured resonances.

\section{Neutral and Charged Exciton Scattering With Electrons}\label{sec:scattering}

A theory of neutral and charged exciton electron scattering
has been proposed by \citet{ramon:2003}. They investigated both
elastic and inelastic scattering in GaAs quantum wells. They were
able to fit their results on the high energy line shape of excitons, but failed to
check their validity on the high energy tail of trions
because in GaAs exciton and trion lines are not
well separated.

We applied their method to calculate the linewidth broadening due to
the exciton-electron scattering in our CdTe quantum well. Here, we
summarize the principles and show the results.

\begin{figure}[tb!]
  \includegraphics[width=0.8\linewidth]{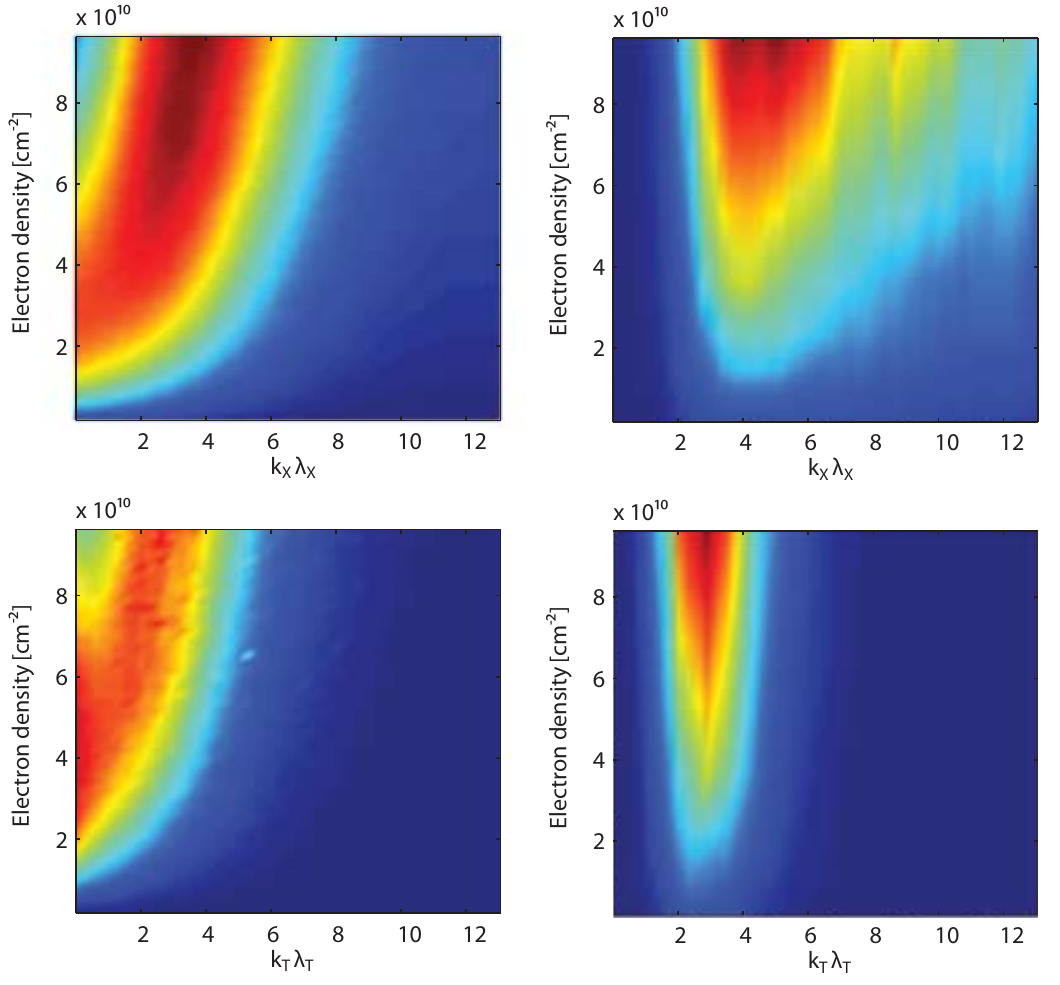}\\
  \caption{Top row: exciton-electron (a) elastic and (b) inelastic scattering. Bottom row: trion-electron (c) elastic and (d) inelastic scattering.}
  \label{fig:scattering}
\end{figure}

\subsection{Exciton-electron scattering}

An electron of momentum $\bm{k}_e$ can interact elastically with an
exciton $\bm{k}_X$. In that case, a momentum $\bm{q}$ is transferred
from the electron to the exciton
\begin{equation}\label{eq:XeScateringPrinciple}
    (\bm{k}_X,\bm{k}_e)\longrightarrow
    (\bm{k}_X+\bm{q},\bm{k}_e-\bm{q}).
\end{equation}
In the process, the electron bound in the exciton can be exchanged
with the free electron. The exchange scattering matrix element is
very strong compared to direct Coulomb interaction. Using Fermi's
golden rule, it is possible to calculate the scattering rate of the
process~\ref{eq:XeScateringPrinciple}. Summing over all final
exciton states, results --- in the first Born approximation --- in
the exciton linewidth $\Gamma_{X-e}^{\text{elastic}}(k_X)$ due to
elastic scattering, as a function of its initial momentum.

Fig.~\ref{fig:scattering}(a) shows the computed values of
$\Gamma_{X-e}^{\text{elastic}}(\bm{k}_X)$ as a function of the
exciton initial energy and the electron density. The large linewidth
obtained for relatively low electron density reflects the high
efficiency of the electron-scattering mechanism. This should be
compared to an exciton linewidth of $\sim 0.1$~meV for acoustic
phonon scattering at T=$5$~K. It is explained by the
exciton-electron interaction matrix elements that favor small
energy-transfer transitions. At higher densities the effect of the
phase-space filling becomes noticeable and effectively enlarges the
exciton Bohr radius. Increasing the electron density further results
in a shift of the maximum linewidth from $k_X=0$ to higher momenta.

During the scattering process, the exciton can also ionized into a
free electron-hole pair. The exciton linewidth
$\Gamma_{X-e}^{\text{inelastic}}(\bm{k}_X)$ due to this inelastic
scattering is represented in Fig.~\ref{fig:scattering}(b). Although
the magnitude of $\Gamma_{X-e}^{\text{inelastic}}(kx)$ is of the
same order as $\Gamma_{X-e}^{\text{elastic}}(kx)$, its functional
dependence on the exciton in-plane momentum is very different. In
particular, we note that the maximal linewidth is obtained at a very
large momentum. This is due to the fact that in order for an exciton
with initially small $k_X$ to be ionized, it must scatter on an
electron with energy large enough to overcome its binding energy. At
zero temperature this is only possible above the Fermi energy.

The two scattering processes that we described above contribute to
admix states with $k_X>0$ to the $k_X=0$ state. This admixture can
be easily included in our previous calculation of the absorption by
convoluting the imaginary part of the exciton
susceptibility~(\ref{eq:chiXT1}) with a Lorentzian function whose
broadening is given by the sum
$\Gamma_{X-e}(k_X)=\Gamma_{X-e}^{\text{elastic}}(k_X)+\Gamma_{X-e}^{\text{inelastic}}(k_X)$
(Fig.~\ref{fig:principle}). It becomes
\begin{multline}\label{eq:ChiWithScattering}
    \text{Im}\left[\chi_{QW}(\omega)\right] = \\ \int_0^\infty \ud{\omega'} \text{Im}\left[\chi_X(\omega-\omega')\right]\mathscr{L}(\omega',\Gamma_{X-e}(\omega'))
\end{multline}
Use of the Kramers-Kronig relations yields the real part of the
dielectric function. Since $\Gamma_{X-e}(k_X)$ is a decreasing
function of $k_X$, the Lorentzian peak is shifted to higher exciton
energies. This is seen as the electron density increases.

\begin{figure}[thb]
    \begin{center}
        \includegraphics[width=0.4\textwidth]{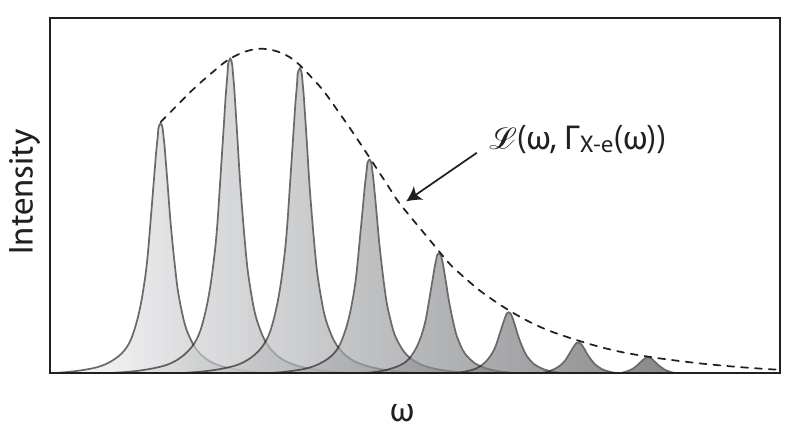}\\
    \end{center}
  \caption{A schematic picture of the convolution resulting from
electron scattering. The solid lines represent various exciton
initial line shapes $\textrm{Im}(\chi_QW(\omega-\omega'))$. The
contributions of states with $k_X>0$ to the $k_X=0$ state are
weighted by the value of the Lorentzian function
$\mathscr{L}_{x-e}[\omega,\Gamma_{x-e}(\omega)]$, given by the
dashed line}\label{fig:principle}
\end{figure}

\subsection{Trion-electron scattering}

For trions, the charge of the trion results in a divergence of its
matrix elements in the limit of zero transferred momentum. This
divergence, originating from the infinite range of the Coulomb
potential, is treated by applying the Lindhard model for the
potential screening. We use the Chandrashekar's variational function
introduced above to perform the calculation. As for excitons, the
broadening comes form direct and exchange scattering. The broadening
$\Gamma_{X^--e}(k_T)$ obtained is shown in Fig.~\ref{fig:principle}(c) and~\ref{fig:principle}(d) .

\section{Electron scattering correction}\label{sec:CWFitScattering}

Using the CdTe exciton-electron and trion-electron scattering
calculation performed in the previous section, we were able to fit
the three cw reflectivity spectra considered in
Fig.~\ref{fig:ReflectivityFit_exchange}. For each resonance
(exciton and trion), the free parameters were the spectral line
position, the oscillator strength, the electron density, the
homogenous and the inhomogeneous linewidth. The fits are shown in
Fig.~\ref{fig:ReflectivityFit_exchange} and the parameters of the
fit in Table~\ref{table:ScatteringParameters}.

\begin{table}[tb]
\begin{center}
\begin{tabular}{l c c c c c }
\hline
&$n_e$ [cm$^{-2}$]&$E$ [meV]&$\Gamma_{\text{hom}}$ [meV]&$\Gamma_{\text{inhom}}$ [meV]\\ \hline  %
$X$ & $4.9\times10^{10}$ & 1623.9 &  0.25 & 0.1\\
$X$ & $6.7\times10^{10}$ & 1624.1 &  0.20 & 0.05\\
$X$ & $1.2\times10^{11}$ & 1624.4 &  0.20 & 0.05\\
$X^-$ & $4.9\times10^{10}$ & 1621.9  & 0.2 & 0.05\\
$X^-$ & $6.7\times10^{10}$ & 1621.9  & 0.3 & 0.05\\
$X^-$ & $1.2\times10^{11}$ & 1622.1  & 0.3 & 0.10\\
\hline
\end{tabular}
\end{center}
 \caption{Fitting parameters} \label{table:ScatteringParameters}
 \end{table}

The fits we obtained are exceptionally good. Only a very small
inhomogenous broadening had to be included, which is consistent with
the high quality of our quantum well. The homogenous broadening
$\gamma_X$ and $\gamma_T$ attributed to phonons remained quite small
and more or less constant over the densities. This shows that we
accounted for almost all electron induced broadening in our simple
model. Had it been perfect, we would have expected the exciton and
trion energy parameters to stay constant for all densities.
Unfortunately, we could not get a perfect match of the spectrum
lineshape by keeping the shift constant.  Our model accounts for
part of the shift but not all. We remind that it is a simple first
order perturbation model and that we might expect higher order
contributions to play a role. The density obtained remained quite
close to those calculated with a simpler model in
Sec.~\ref{sec:experiment}.

As for the oscillator strength, not much can be said, since we did
not include any dependence on the electron density in our model. We
did so intentionally because current theoretical models conclude
that the trion oscillator strength depends on the volume of the
sample~\citep{esser:2002}. We think that it is a failure of the
three-particle oversimplified trion model which is, strictly
speaking, only correct in quantum dots. A correct description of the
trion resonance should take into account the interaction of the
excitons with all electron in the system. Such approach seems to
lead to trion oscillator strength that do not depend on the
volume~\citep{dupertuis:2006}.

\section{Time-resolved reflectivity experiment}\label{sec:tr-experiment}

Spectrally resolved pump and probe experiments were also performed in reflectivity geometry at 5 K in order to investigate the effect of the exciton-electron and trion-electron scattering on the nonlinear reflectivity spectra. We use the sample without illumination so with an excess electron population of about $4\times10^{10}$~cm$^{-2}$ in the quantum well. The 100 femtosecond output pulses from of a Ti:Saphire laser were split into two: a small portion circulated through a delay line and probed the exciton and trion resonances, while the major portion passed through a pulse shaper to generate 4 ps long and 0.5 meV wide pump pulses, with sufficient spectral resolution to selectively pump the trion resonance. The pump/probe intensity ratio was larger than 50 so that probe reflectivity spectra remained linear in the probe field. The polarization of both pump and probe pulses could be chosen independently to investigate the counter-circular $\sigma^+\sigma^-$ polarization configuration. We recorded the differential reflectivity spectrum $\Delta R=(R-R_0)/R_0$ for a given delay time between pump and probe pulses, $R_0$ and $R$ being the unexcited and excited reflectivity spectrum, respectively.
 
 We choose the excitation pulse to be at trion resonance, because by generating trions we induce inevitably a variation in the electron population. We decide to perform the experiments with $\sigma^+\sigma^-$ circular polarization because with this configuration we avoid phase space filling effects [34] enabling us to detect changing of the exciton as well the trion resonance lineshapes. 
  In Figure 5, we show the differential reflectivity spectra obtained by pumping at trion resonance and probing with at delay time of 4 ps. At this short positive delay time, all coherent processes have decayed away \cite{berney:2007}, being it then the best delay to observe nonlinearities on the exciton and trion resonances. Therefore, we can in this way check the changing in the exciton and trion resonances due to electron scattering.
  
Indeed, the differential reflectivity spectrum in Fig. 5 evidences variations in both trion and exciton resonances. We observe that the trion line is slightly redshifted. We attribute this redshift to the reduced amount of free electrons in the quantum well by the $\sigma^+$ trion generation. This induces a decrease scattering contributions from the electrons with the probed $\sigma^-$ trions. Therefore, the trion high energy tail absorption will reduce and as a consequence, its lineshape will change shifting a little towards lower energies. The trion generation also brings changing in the exciton resonance. We observe the expected bleaching of the high energy tail of the exciton absorption with an induced absorption of the exciton resonance. With these results, by generating a trion population and therefore altering the electron distribution, we evidence the importance of electron scattering with excitons and trions in their lineshapes and in their nonlinear signal. 

\begin{figure}[thb]
    \begin{center}
        \includegraphics[width=0.3\textwidth]{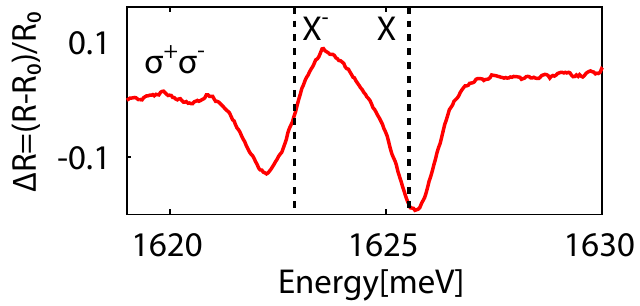}\\
    \end{center}
  \caption{Differential reflectivity spectra at delay time of 4 ps for  $\sigma^+\sigma^-$ pump and probe configurations. The pump $\sigma^+$ is at trion resonance.}\label{fig:principle}
\end{figure}

\section{Conclusion}\label{sec:conclusion}
   
In summary, we demonstrated that in modulation doped quantum wells, electrons play an important role on both exciton and trion resonance lineshapes. We evidenced experimentally that a high energy tail develops at both excitons and trions resonances as the excess electron population is increased in the quantum well. We showed that with a model without considering exciton and trion-electron scatterings, we are not able to reproduce the optical spectra. We used then a simple first order perturbation model taking into account elastic and inelastic scatterings of electrons with excitons and trions. The model is based on calculating the exciton - and trion-electron direct and exchange interaction matrix elements to derive the trion and the exciton linewidth broadening. Finally, by including theses results in the matrix transfer calculations, we reproduced the highly asymmetrical resonance shapes of excitons and trions in the reflectivity spectra. This is fundamental for the interpretation of the linear and non-linear properties of modulation-doped quantum wells.
  
Acknowledgments: we acknowledge financial support from FNRS within quantum photonics NCCR. We thank C. Ciuti, M. A. Dupertuis and V. Savona, for enlightening discussions as well as M. Kutrowski and T. Wojtowicz for growing our sample. 

\appendix

\section{Matrix Transfer Formalism for a Stratified Medium} \label{appendix:TM}

The layered structure of our sample needs to be taken into account
in order to consistently simulate the optical properties of the
quantum well. We study the propagation of time-harmonic electromagnetic wave through a stratified medium comprising successive thin plane-parallel films.

First we shall solve the wave equation
\begin{equation}\label{eq:MaxwellElectric}
    \bnabla^2\bm{E}(\omega,\bm{r}_\parallel,z) - \frac{\omega^2}{\epsilon_0 c^2}
    \bm{D}(\omega,\bm{r}_\parallel,z)=0.
\end{equation}
where the electric field $\bm{E}(\omega,\bm{r}_\parallel,z)$ and
the electric displacement $\bm{D}(\omega,\bm{r}_\parallel,z)$ are expressed
in term of the frequency $\omega$, the in-plane position vector
$\bm{r}_\parallel$ and the position  along the growth axis $z$.

For a homogenous dielectric medium, the electric and displacement
fields are linearly dependent $\bm{D}(\omega,\bm{r}_\parallel,z)
=\epsilon(\omega)\bm{E}(\omega,\bm{r}_\parallel,z)$.
Eq.~(\ref{eq:MaxwellElectric}) takes the form of a Helmholtz wave
equation for $\bm{E}(\omega,\bm{r}_\parallel,z)$
\begin{equation}\label{eq:Helmoltz}
    \left(\bnabla^2 + \frac{\omega^2}{c^2}\frac{\epsilon(\omega)}{\epsilon_0} \right)\bm{E}(\omega,\bm{r}_\parallel,z) =
    0. 
\end{equation}
The dielectric dispersion $\epsilon(\omega)$ may very well be
complex. Because it is constant, the electromagnetic field is
invariant under in-plane translations (Bloch theorem) and solutions
of (\ref{eq:Helmoltz}) are plane-waves
\begin{equation}\label{eq:ElectricField}
    \bm{E}_{\bm{k}_\parallel}(\omega,\bm{r}_\parallel,z) = \bm{\epsilon}_{\bm{k}_\parallel}u_{\bm{k}_\parallel}(\omega,z) \ee^{i
    \bm{k}_\parallel\cdot\bm{r}_\parallel}.
\end{equation}
Here $\bm{k}_\parallel$ is the in-plane wave vector and
$\bm{\epsilon}_{\bm{k}_\parallel}$ the polarization vector. After
substitution into (\ref{eq:Helmoltz}), we are left with a
one-dimensional problem for the mode function
$u_{\bm{k}_\parallel}(\omega,z)$
\begin{equation}\label{eq:Helmoltz1D}
    \frac{\dd^2 u_{\bm{k}_\parallel}(\omega,z)}{\dd
    z^2}+\left(\frac{\omega^2}{c^2}\frac{\epsilon(\omega)}{\epsilon_0}-k^2_{\parallel}\right)u(\omega,z)=0.
\end{equation}
The solution represents two counter-propagating waves
\begin{gather}
    \label{eq:ElectricField1D}%
    u(\omega,\bm{k}_\parallel,z)
        = E_l(\bm{k}_\parallel)\ee^{-ik_zz}+E_r(\bm{k}_\parallel)\ee^{ik_zz},\\
    \label{eq:kz}%
    k_z =
    \sqrt{\frac{\omega^2}{c^2}\frac{\epsilon(\omega)}{\epsilon_0}-k_\parallel}.
\end{gather}
The coefficient $E_l$ and $E_r$ are two complex coefficient that
have to be determined by imposing Maxwell boundary conditions at the
interface between adjacent layers. Let us note that if
$\epsilon(\omega)$ is imaginary then $k_z$ is imaginary as well and
the solutions are evanescent waves.

We use transfer matrices to express the propagation of the electromagnetic wave at the interface between two consecutive films with refraction index $n_1$ and $n_2$. Then, we combine the matrices to calculate analytically the mode of the whole stratified medium.

\begin{figure}[thb]
    \begin{center}
        \includegraphics[width=0.3\textwidth]{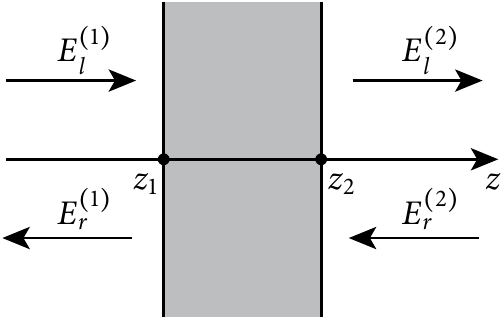}\\
    \end{center}
  \caption{Electric Fields at the boundary $z_1$ and $z2$ of a plane layer.}\label{fig:TM}
\end{figure}

In the framework of the one-dimensional
problem~(\ref{eq:Helmoltz1D}) we define for each position~$z$ in
space a two-dimensional vector containing the two coefficients
in~(\ref{eq:ElectricField1D})
\begin{equation}\label{eq:TransferMatrixVector}
    \bm{Q}=
    \begin{bmatrix}
      E_r\\
      E_l\\
    \end{bmatrix}.
\end{equation}
For a given structure, we would like to express the relation between
coefficient vectors $\bm{Q}^{(1)}$ at~$z_1$
and $\bm{Q}^{(2)}$ at~$z_2$.
Because Maxwell equations are linear, Maxwell boundary conditions
will result in a linear relation, which we write as
\begin{equation}\label{eq:TransferMatrixDefinition}
\bm{Q}^{(2)}=\bm{M}\bm{Q}^{(1)}.
\end{equation}
The $2\times2$ complex matrix $\bm{M}$ thus defined is called the
\emph{transfer} or \emph{propagation matrix}.  If $z_1$ and~$z_2$ were to be found in the same homogenous layer,
from (\ref{eq:ElectricField1D}) the transfer matrix would simply be
\begin{equation}\label{eq:Mhomogenous}
    \bm{M}_{\textrm{hom}}=
    \begin{bmatrix}
      \ee^{ik_z(z_2-z_1)} & 0 \\
      0 & \ee^{-ik_z(z_2-z_1)} \\
    \end{bmatrix}.
\end{equation}

The Maxwell boundary conditions are invariant under time reversal.
This means that the complex coefficient of a transfer matrix $M$ do
not change if we reverse the time evolution~\citep{cargese:1999}.
Mathematically, we write this as
\begin{equation}
    \bm{M}=\hat{T}\bm{M}\hat{T}^{-1},
\end{equation}
where $\hat{T}$ is the time reversal operator. We would like to
calculate $\hat{T}\bm{Q}$. We remind that the amplitude of the
electric field is given by the real part of its representation in
terms of complex exponentials
\begin{equation*}\label{eq:ElectricField}
E(\omega,\bm{k}_\parallel,\bm{r}_\parallel,z,t)
=\Re\left[\left(E_l\ee^{-ik_zz}+E_r\ee^{ik_zz}\right)\ee^{i\bm{k}_\parallel\bcdot\bm{r}_\parallel}\ee^{-i\omega{t}}\right].
\end{equation*}
The time reversal operator then acts as
\begin{multline*}\label{eq:ElectricFieldT}
\hat{T}E(\omega,\bm{k}_\parallel,\bm{r}_\parallel,z,t)
=E(\omega,\bm{k}_\parallel,\bm{r}_\parallel,z,-t)\\
=\Re\left[\left(E_r^*\ee^{-ik_zz}+E_l^*\ee^{ik_zz}\right)\ee^{-i\bm{k}_\parallel\bcdot\bm{r}_\parallel}\ee^{-i\omega{t}}\right].
\end{multline*}
Thus, $\widehat{T}$ reverse the sign of
$\bm{k}_\parallel\rightarrow-\bm{k}_\parallel$ and acts on $\bm{Q}$
as
\begin{equation}\label{eq:TransferMatrixDefinition}
    \widehat{T}
    \begin{bmatrix}
      E_r \\
      E_l \\
    \end{bmatrix}
    =
    \begin{bmatrix}
      E_l^* \\
      E_r^* \\
    \end{bmatrix}.
\end{equation}
The time reversal invariance let us express the transfer matrix of a
stratified medium in terms of its complex reflectivity and transfer
coefficients. We consider the case of a unitary wave which is
arriving on the left, reflected in the opposite direction with a
wave amplitude $r$ and transmitted to the right with an amplitude
$t$. Then
\begin{equation}\label{eq:TransferMatrixRTnor}
    \begin{bmatrix}
      ^{\phantom{*}}t^{\phantom{*}} \\
      0 \\
    \end{bmatrix}
    =
    \begin{bmatrix}
      M_{11} & M_{12} \\
      M_{21} & M_{22} \\
    \end{bmatrix}
    \begin{bmatrix}
      1 \\
      ^{\phantom{*}}r^{\phantom{*}} \\
    \end{bmatrix}
\end{equation}
By applying the time reversal operator $\hat{T}$ on both side of the
equality we get a second system of equations
\begin{equation}\label{eq:TransferMatrixRTrev}
    \begin{bmatrix}
      0 \\
      ^{\phantom{*}}t^* \\
    \end{bmatrix}
    =
    \begin{bmatrix}
      M_{11} & M_{12} \\
      M_{21} & M_{22} \\
    \end{bmatrix}
    \begin{bmatrix}
      ^{\phantom{*}}r^* \\
      1 \\
    \end{bmatrix}.
\end{equation}

Noting that in terms of $r$ and $t$, the reflectivity and
transmissivity are $R=|r|^2$ and $T=\alpha_{12}^{-1}|t|^2$ with
\begin{equation}
    \alpha_{12}=
    \begin{cases}
        {\Re\left[k_{z}^{(1)}\right]} / {\Re\left[k_{z}^{(2)}\right]} & \text{for TE polarization}\\
        {\Re\left[k_{z}^{(1)}\right]}n_2^2 /
        {\Re\left[k_{z}^{(2)}\right]n_1^2} & \text{for TM polarization}
    \end{cases},
\end{equation}
we solve the system of linear equations
(\ref{eq:TransferMatrixRTnor}) and (\ref{eq:TransferMatrixRTrev})
and obtain
\begin{equation}\label{eq:TransferMatrixRT}
    M = \alpha_{12}^2
    \begin{bmatrix}
      \cfrac{1}{t^*} & -\cfrac{r^*}{t^*} \\
      -\cfrac{r}{t} & \cfrac{1}{t} \\
    \end{bmatrix}.
\end{equation}

Then, we use Eq.~(\ref{eq:TransferMatrixRT}) here above to calculate
the transfer matrix at an interface between two homogenous
layers~\citep{born:1999}.
We obtain the two transfer matrices for TE and TM polarization
\begin{align}\label{eq:TransferMatrixTE}%
    \bm{M}^\textrm{TE}&= \alpha_{12}^{\textrm{TE}}
    \begin{bmatrix}
      \cfrac{k_z^{(1)}+k_z^{(2)}}{2k_z^{(1)}} & \cfrac{k_z^{(1)}-k_z^{(2)}}{2k_z^{(1)}} \\
      \cfrac{k_z^{(1)}-k_z^{(2)}}{2k_z^{(1)}} & \cfrac{k_z^{(1)}+k_z^{(2)}}{2k_z^{(1)}} \\
    \end{bmatrix}
\end{align}
and
\begin{align}
\label{eq:TransferMatrixTM}%
    \bm{M}^\textrm{TM}&= \alpha_{12}^{\textrm{TM}}
    \begin{bmatrix}
      \cfrac{\epsilon^{(2)}k_z^{(1)}+\epsilon^{(1)}k_z^{(2)}}{2\sqrt{\epsilon^{(1)}\epsilon^{(2)}}k_z^{(1)}}
      & \cfrac{\epsilon^{(2)}k_z^{(1)}-\epsilon^{(1)}k_z^{(2)}}{2\sqrt{\epsilon^{(1)}\epsilon^{(2)}}k_z^{(1)}} \\
      \cfrac{\epsilon^{(2)}k_z^{(1)}-\epsilon^{(1)}k_z^{(2)}}{2\sqrt{\epsilon^{(1)}\epsilon^{(2)}}k_z^{(1)}}
      & \cfrac{\epsilon^{(2)}k_z^{(1)}+\epsilon^{(1)}k_z^{(2)}}{2\sqrt{\epsilon^{(1)}\epsilon^{(2)}}k_z^{(1)}} \\
    \end{bmatrix}.
\end{align}
These expression are quite general and are valid for complex
dielectric constants as well. Combining those two matrices with the
propagation matrix~(\ref{eq:Mhomogenous}) we are able to calculate
the electromagnetic modes in any multilayered structure.

\section{Transfer Matrix of a Quantum Well}\label{fig:TMQW}

We solve Eq.~\ref{eq:MaxwellElectric} for a quantum
well of with width $L$ when TE polarized light propagates in the
barrier. The electric field is along the $y$-axis and
(\ref{eq:MaxwellElectric}) becomes
\begin{equation}\label{eq:HelmoltzQWTE}
    (\partial_z^2+k_z^2)E_y+k^2\chi(\omega)\rho(z)\int_{-L/2}^{L/2}\ud{z'}\rho(z')E_y(z')=0
\end{equation}
with $k=\sqrt{{\epsilon(\omega)}/{\epsilon_0}}\,{\omega}/{c}$. As
$\rho(z)$ is even for optical allowed transitions, it is convenient
to treat separately solutions of the problem with definite parity
under inversion of the $z$ coordinate.

For odd $E_y$ symmetry, the integral in \ref{eq:HelmoltzQWTE} is
zero and there is no polarization contribution from the quantum
well. The reflectivity takes the simple form
\begin{equation}\label{eq:OddReflectivtyTE}
    r^{\text{(odd)}}_{\text{TE}}=-e^{ik_zL}.
\end{equation}

For even $E_y$ symmetry, the solution of the second order
differential equation (\ref{eq:HelmoltzQWTE}) is
\begin{equation}\label{eq:HelmoltzQWTESolution}
    E_y=A(\omega,k_z)\cos(k_zz) + \int \ud{z'}G(z,z')\rho(z').
\end{equation}
In this expression, the first term is the general even solution of
the problem $(\partial_z^2+k_z^2)E_y=0$, with $A(\omega,k_z)$ a
constant to be determined. The second term is a particular solution
of (\ref{eq:HelmoltzQWTE}) and was simply built from the definition
of the Green function $(\partial_z^2+k_z^2)G(z,z')=\delta(z-z')$. We
choose
\begin{equation}\label{eq:GreenFunction}
    G(z,z')=-\frac{1}{2k_z}\sin(k_z|z-z'|)
\end{equation}
which has the advantage to yield an even solution of $E_y$.
Replacing (\ref{eq:HelmoltzQWTESolution}) into
(\ref{eq:HelmoltzQWTE}) gives
\begin{align}\label{eq:EvenReflectivityTE}
    A(\omega,k_z)&=(\chi(\omega)^{-1}-P(k_z)/Q(k_z)\\
\label{eq:Qkz}
    Q(k_z)&=\int_{-L/2}^{L/2}\ud{z}\rho(z)\cos(k_zz)\\
\label{eq:Pkz}
    P(k_z)&=\int_{-L/2}^{L/2}\int_{-L/2}^{L/2}\ud{z}\ud{z'}\frac{1}{2k_z}\sin(k_z|z-z'|)\rho(z)\rho(z')
\end{align}
We consider Maxwell boundary condition at the interface $z=-L/2$. On
the left side of the quantum well, we know that there are two
counter-propagating plane waves propagating in the barrier given by
Eqs.~\ref{eq:ElectricField} and \ref{eq:ElectricField1D}. The
Maxwell boundary conditions give
\begin{multline}\label{eq:MaxwellBoundaryQW}
    (E_y^{(l)}+E_y^{(r)})\big|_{z=-L/2^-}=E_y\big|_{z=-L/2^+}\\
    \shoveleft{\partial_z(E_y^{(l)}+E_y^{(r)})\big|_{z=-L/2^-}}\\
    \qquad=ik_z(E_y^{(l)}-E_y^{(r)})\big|_{z=-L/2^-}=\partial_zE_y\big|_{z=-L/2^+}
\end{multline}
Substituing Eq.~\ref{eq:HelmoltzQWTESolution} in those two equation,
we obtain the reflectivity for the even solution:
\begin{align}\label{eq:EvenReflectivityTE}
    r^{\textrm{(even)}}_{\text{TE}}
    &=\frac{E_y^{(r)}}{E_y^{(l)}}
    =\frac{E_y-\frac{1}{ik_z}\partial{E_y}}{E_y+\frac{1}{ik_z}\partial{E_y}}\bigg|_{z=-L/2^+}\notag\\
    &=e^{ik_zL}\frac{A(\omega,k_z)-{iQ(k_z)}/{2k_z}}{A(\omega,k_z)+{iQ(k_z)}/{2k_z}}\notag\\
    &=e^{ik_zL}\frac{1-\chi(\omega,k_z)(i\alpha(k_z)+P(\omega,k_z))}{1+\chi(\omega,k_z)(i\alpha(k_z)-P(\omega,k_z))}
\end{align}
where $\alpha(k_z)=Q(k_z)^2/2k_z$. The total reflectivity is given
by the mean of the odd~\ref{eq:OddReflectivtyTE} and even
reflectivity~\ref{eq:EvenReflectivityTE}:
\begin{align}
    r_{\text{TE}}^{\phantom{()}}
    &=\tfrac{1}{2}\left(r_{\text{TE}}^{\textrm{(even)}}+r_{\text{TE}}^{\textrm{(odd)}}\right)\notag\\
    &=\frac{-i\alpha(k_z)\chi(\omega,k_z)e^{ik_zL}}{1+\chi(\omega,k_z)(i\alpha(k_z)-P(\omega,k_z))}
\end{align}
The same kind of calculation apply for the transmissivity, which
reads
\begin{align}
    t_{\text{TE}}^{\phantom{()}}&=\tfrac{1}{2}\left(t_{\text{TE}}^{\textrm{(even)}}-t_{\text{TE}}^{\textrm{(odd)}}\right)\notag\\
    &=\frac{(1-P(\omega,k_z)\chi(\omega,k_z))e^{ik_zL}}{1+\big[i\alpha(k_z)-P(\omega,k_z)\big]\chi(\omega,k_z)}
\end{align}
If we now suppose that the quantum well has an infinite potential,
the calculation of $\alpha$ and $P$ functions becomes trivial. We
obtain $\alpha\rightarrow 1$ and $P\rightarrow 0$. We are left with
the simple reflectivity and transitivity
\begin{align}
    r_{\text{TE}}^{\phantom{()}}
    =e^{ik_zL}\frac{-i\chi(\omega,k_z)/2k_z}{1+i\chi(\omega,k_z)/2k_z}
\end{align}
\begin{align}
    t_{\text{TE}}^{\phantom{()}}
    =e^{ik_zL}\frac{1}{1+i\chi(\omega,k_z)/2k_z}=1+r_{\text{TE}}^{\phantom{()}}
\end{align}
When used in Eq.~\ref{eq:TransferMatrixRT}, these two expressions
give the transfer matrix of a quantum well. The only information
that is required to simulate the linear optical spectrum of our
quantum well is the susceptibility.

\section{Variational Wavefunctions}\label{appendix:wf}

For the trion wavefunction, we relied on a simpler variational
function \citep{sergeev:2001}, given by
\begin{multline}\label{eq:sergeev}
\psi_T(\bm{\rho_1},\bm{\rho_2})=(e^{-a \rho_1 -b \rho_2} + e^{-a
\rho_2 -b \rho_1})\\ \times(1+c R)  \frac{e^{-s R}}{1+d(R-R_0)}
\end{multline}
where $\bm{R}=\bm{r}_{1e}-\bm{r}_{2e}$. The variational parameters
$a$, $b$, $c$, $d$, $s$, $R_0$ were obtained by fitting the
numerical solution of the Schr{\"o}dinger
equation~\citep{esser:2000}. They are given in
Table~\ref{table:sergeev} for both positive and negative trion in
units of the bulk CdTe bulk Bohr radius ($7.75$~nm). The overlap
with the numerical wave function is excellent ($0.9981$ for $X^-$).

\begin{table}[t]
\begin{center}
\begin{tabular}{c c c c c c c }
\hline
&$a$&$b$&$c$&$d$&$s$&$R_0$\\ \hline  %
$X^-$ & 1.1723 & 0.5471 & 0.4809 & 0.0682 & 0.0013 & 1.1614\\
$X^+$ & 1.1774 & 0.4294 & 2.3943 & 0.08431 & 0.26173 & 0.7393\\
\hline
\end{tabular}
\end{center}
 \caption{Variational parameters for Sergeev and Suris' trion variational function} \label{table:sergeev}
 \end{table}

\begin{figure}[!b]
    \begin{center}
        \includegraphics[width=\linewidth]{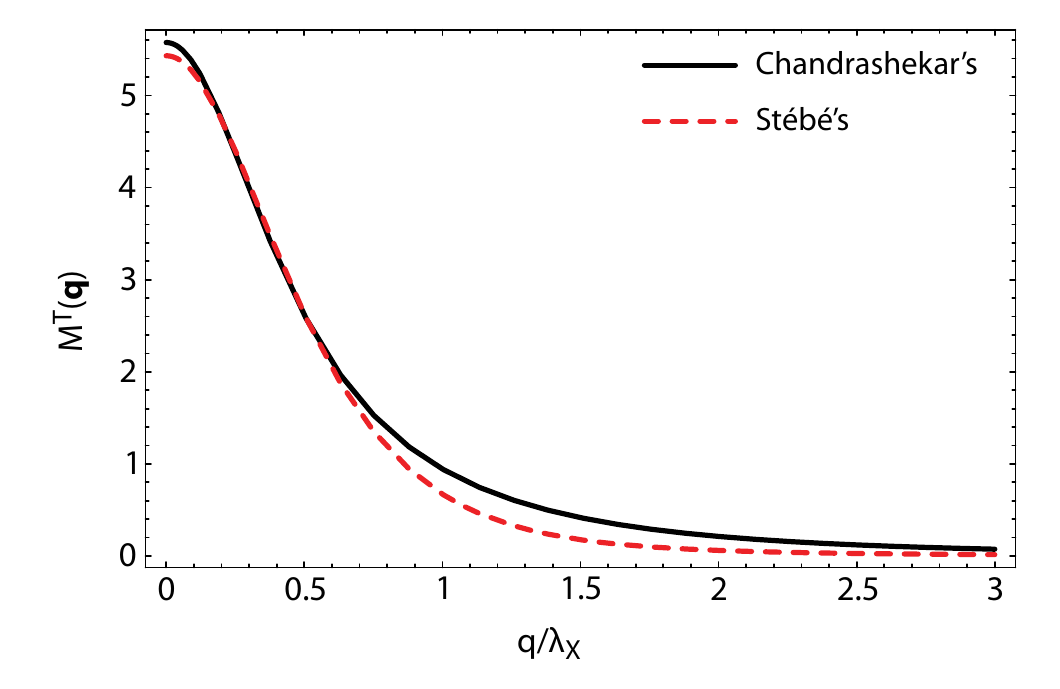}
    \end{center}
\caption[Trion optical coupling element.]{Trion optical coupling as
a function of the momentum $\bm{q}$ of the initial electron.}
\label{fig:Mtrion}
\end{figure}

Even if this variational function is analytical, its Fourier
transform is not. To speed up calculations, we resorted to a simpler
Chandrashekar's wave function \citep{chandrasekhar:1944} with only
two variational parameters $\lambda=6.8$~nm and $\lambda'=15$~nm. It
is given by
\begin{equation}\label{eq:chandrasekhar}
\psi_T(\rho_1,\rho_2)= \mathcal{N}_T \left(e^{\rho_1/\lambda
-\rho_2/\lambda'} + e^{\rho_1/\lambda' -\rho_2/\lambda} \right),
\end{equation}
Where $\mathcal{N}_T$ is the normalization factor. With this
approximation, the light coupling element $M(\bm{q})$
 becomes
\begin{equation}\label{eq:MT}
    M_T(\bm{q})=4 \mathcal{N}_T\left(\frac{\lambda_T}{\lambda_T'}\frac{1}{(1+(\lambda_T
    \bm{k})^2)^{3/2}}+\lambda_T\leftrightarrow\lambda_T'\right).
\end{equation}

We evaluated the error induce by using the Chandrashekar's
wavefunction. In Fig.~\ref{fig:Mtrion}, we compare the result
obtained with variational functions \ref{eq:sergeev} and
\ref{eq:chandrasekhar}. They do not differ much and we conclude that
Eq.~\ref{eq:MT} is a reliable approximation.


\begin{thebibliography}{31}
\expandafter\ifx\csname natexlab\endcsname\relax\def\natexlab#1{#1}\fi
\expandafter\ifx\csname bibnamefont\endcsname\relax
  \def\bibnamefont#1{#1}\fi
\expandafter\ifx\csname bibfnamefont\endcsname\relax
  \def\bibfnamefont#1{#1}\fi
\expandafter\ifx\csname citenamefont\endcsname\relax
  \def\citenamefont#1{#1}\fi
\expandafter\ifx\csname url\endcsname\relax
  \def\url#1{\texttt{#1}}\fi
\expandafter\ifx\csname urlprefix\endcsname\relax\def\urlprefix{URL }\fi
\providecommand{\bibinfo}[2]{#2}
\providecommand{\eprint}[2][]{\url{#2}}

\bibitem[{\citenamefont{Chemla and Miller}(1985)}]{chemla:1985}
\bibinfo{author}{\bibfnamefont{D.~S.} \bibnamefont{Chemla}} \bibnamefont{and}
  \bibinfo{author}{\bibfnamefont{D.~A.~B.} \bibnamefont{Miller}},
  \bibinfo{journal}{Journal of the Optical Society of America B}
  \textbf{\bibinfo{volume}{2}}, \bibinfo{pages}{1155} (\bibinfo{year}{1985}).

\bibitem[{\citenamefont{Huard et~al.}(2000)\citenamefont{Huard, Cox,
  Saminadayar, Arnoult, and Tatarenko}}]{huard:2000}
\bibinfo{author}{\bibfnamefont{V.}~\bibnamefont{Huard}},
  \bibinfo{author}{\bibfnamefont{R.~T.} \bibnamefont{Cox}},
  \bibinfo{author}{\bibfnamefont{K.}~\bibnamefont{Saminadayar}},
  \bibinfo{author}{\bibfnamefont{A.}~\bibnamefont{Arnoult}}, \bibnamefont{and}
  \bibinfo{author}{\bibfnamefont{S.}~\bibnamefont{Tatarenko}},
  \bibinfo{journal}{Physical Review Letters} \textbf{\bibinfo{volume}{84}},
  \bibinfo{pages}{187} (\bibinfo{year}{2000}).

\bibitem[{\citenamefont{Kossacki et~al.}(1999)\citenamefont{Kossacki, Cibert,
  Ferrand, d'Aubigne, Arnoult, Wasiela, Tatarenko, and Gaj}}]{kossacki:1999}
\bibinfo{author}{\bibfnamefont{P.}~\bibnamefont{Kossacki}},
  \bibinfo{author}{\bibfnamefont{J.}~\bibnamefont{Cibert}},
  \bibinfo{author}{\bibfnamefont{D.}~\bibnamefont{Ferrand}},
  \bibinfo{author}{\bibfnamefont{Y.~M.} \bibnamefont{d'Aubigne}},
  \bibinfo{author}{\bibfnamefont{A.}~\bibnamefont{Arnoult}},
  \bibinfo{author}{\bibfnamefont{A.}~\bibnamefont{Wasiela}},
  \bibinfo{author}{\bibfnamefont{S.}~\bibnamefont{Tatarenko}},
  \bibnamefont{and} \bibinfo{author}{\bibfnamefont{J.~A.} \bibnamefont{Gaj}},
  \bibinfo{journal}{Physical Review B} \textbf{\bibinfo{volume}{60}},
  \bibinfo{pages}{16018} (\bibinfo{year}{1999}).

\bibitem[{\citenamefont{Yusa et~al.}(2000)\citenamefont{Yusa, Shtrikman, and
  Bar-Joseph}}]{yusa:2000}
\bibinfo{author}{\bibfnamefont{G.}~\bibnamefont{Yusa}},
  \bibinfo{author}{\bibfnamefont{H.}~\bibnamefont{Shtrikman}},
  \bibnamefont{and}
  \bibinfo{author}{\bibfnamefont{I.}~\bibnamefont{Bar-Joseph}},
  \bibinfo{journal}{Physical Review B} \textbf{\bibinfo{volume}{62}},
  \bibinfo{pages}{15390} (\bibinfo{year}{2000}).

\bibitem[{\citenamefont{Schultheis et~al.}(1986)\citenamefont{Schultheis, Kuhl,
  Honold, and Tu}}]{schultheis:1986}
\bibinfo{author}{\bibfnamefont{L.}~\bibnamefont{Schultheis}},
  \bibinfo{author}{\bibfnamefont{J.}~\bibnamefont{Kuhl}},
  \bibinfo{author}{\bibfnamefont{A.}~\bibnamefont{Honold}}, \bibnamefont{and}
  \bibinfo{author}{\bibfnamefont{C.~W.} \bibnamefont{Tu}},
  \bibinfo{journal}{Physical Review Letters} \textbf{\bibinfo{volume}{57}},
  \bibinfo{pages}{1635} (\bibinfo{year}{1986}).

\bibitem[{\citenamefont{Honold et~al.}(1989)\citenamefont{Honold, Schultheis,
  Kuhl, and Tu}}]{honold:1989}
\bibinfo{author}{\bibfnamefont{A.}~\bibnamefont{Honold}},
  \bibinfo{author}{\bibfnamefont{L.}~\bibnamefont{Schultheis}},
  \bibinfo{author}{\bibfnamefont{J.}~\bibnamefont{Kuhl}}, \bibnamefont{and}
  \bibinfo{author}{\bibfnamefont{C.~W.} \bibnamefont{Tu}},
  \bibinfo{journal}{Physical Review B} \textbf{\bibinfo{volume}{40}},
  \bibinfo{pages}{6442} (\bibinfo{year}{1989}).

\bibitem[{\citenamefont{Capozzi et~al.}(1993)\citenamefont{Capozzi, Pavesi, and
  Staehli}}]{capozzi:1993}
\bibinfo{author}{\bibfnamefont{V.}~\bibnamefont{Capozzi}},
  \bibinfo{author}{\bibfnamefont{L.}~\bibnamefont{Pavesi}}, \bibnamefont{and}
  \bibinfo{author}{\bibfnamefont{J.~L.} \bibnamefont{Staehli}},
  \bibinfo{journal}{Physical Review B} \textbf{\bibinfo{volume}{47}},
  \bibinfo{pages}{6340} (\bibinfo{year}{1993}).

\bibitem[{\citenamefont{Kheng et~al.}(1993)\citenamefont{Kheng, Cox, d\char39{}
  Aubign\'e, Bassani, Saminadayar, and Tatarenko}}]{kheng:1993}
\bibinfo{author}{\bibfnamefont{K.}~\bibnamefont{Kheng}},
  \bibinfo{author}{\bibfnamefont{R.~T.} \bibnamefont{Cox}},
  \bibinfo{author}{\bibfnamefont{M.~Y.} \bibnamefont{d\char39{} Aubign\'e}},
  \bibinfo{author}{\bibfnamefont{F.}~\bibnamefont{Bassani}},
  \bibinfo{author}{\bibfnamefont{K.}~\bibnamefont{Saminadayar}},
  \bibnamefont{and}
  \bibinfo{author}{\bibfnamefont{S.}~\bibnamefont{Tatarenko}},
  \bibinfo{journal}{PRL} \textbf{\bibinfo{volume}{71}}, \bibinfo{pages}{1752}
  (\bibinfo{year}{1993}).

\bibitem[{\citenamefont{Stebe and Stauffer}(1989)}]{stebe:1989}
\bibinfo{author}{\bibfnamefont{B.}~\bibnamefont{Stebe}} \bibnamefont{and}
  \bibinfo{author}{\bibfnamefont{L.}~\bibnamefont{Stauffer}},
  \bibinfo{journal}{Superlattices and Microstructures}
  \textbf{\bibinfo{volume}{5}}, \bibinfo{pages}{451} (\bibinfo{year}{1989}).

\bibitem[{\citenamefont{Thilagam}(1997)}]{thilagam:1997}
\bibinfo{author}{\bibfnamefont{A.}~\bibnamefont{Thilagam}},
  \bibinfo{journal}{Physical Review B} \textbf{\bibinfo{volume}{55}},
  \bibinfo{pages}{7804} (\bibinfo{year}{1997}).

\bibitem[{\citenamefont{Stebe et~al.}(1997)\citenamefont{Stebe, Munschy,
  Stauffer, Dujardin, and Murat}}]{stebe:1997}
\bibinfo{author}{\bibfnamefont{B.}~\bibnamefont{Stebe}},
  \bibinfo{author}{\bibfnamefont{G.}~\bibnamefont{Munschy}},
  \bibinfo{author}{\bibfnamefont{L.}~\bibnamefont{Stauffer}},
  \bibinfo{author}{\bibfnamefont{F.}~\bibnamefont{Dujardin}}, \bibnamefont{and}
  \bibinfo{author}{\bibfnamefont{J.}~\bibnamefont{Murat}},
  \bibinfo{journal}{Physical Review B} \textbf{\bibinfo{volume}{56}},
  \bibinfo{pages}{12454} (\bibinfo{year}{1997}).

\bibitem[{\citenamefont{Riva et~al.}(2000)\citenamefont{Riva, Peeters, and
  Varga}}]{riva:2000}
\bibinfo{author}{\bibfnamefont{C.}~\bibnamefont{Riva}},
  \bibinfo{author}{\bibfnamefont{F.~M.} \bibnamefont{Peeters}},
  \bibnamefont{and} \bibinfo{author}{\bibfnamefont{K.}~\bibnamefont{Varga}},
  \bibinfo{journal}{Physical Review B} \textbf{\bibinfo{volume}{61}},
  \bibinfo{pages}{13873} (\bibinfo{year}{2000}).

\bibitem[{\citenamefont{Esser et~al.}(2001)\citenamefont{Esser, Zimmermann, and
  Runge}}]{esser:2001}
\bibinfo{author}{\bibfnamefont{A.}~\bibnamefont{Esser}},
  \bibinfo{author}{\bibfnamefont{R.}~\bibnamefont{Zimmermann}},
  \bibnamefont{and} \bibinfo{author}{\bibfnamefont{E.}~\bibnamefont{Runge}},
  \bibinfo{journal}{Physica Status Solidi B} \textbf{\bibinfo{volume}{227}},
  \bibinfo{pages}{317} (\bibinfo{year}{2001}).

\bibitem[{\citenamefont{Suris et~al.}(2001)\citenamefont{Suris, Kochereshko,
  Astakhov, Yakovlev, Ossau, Nurnberger, Faschinger, Landwehr, Wojtowicz,
  Karczewski et~al.}}]{suris:2001}
\bibinfo{author}{\bibfnamefont{R.~A.} \bibnamefont{Suris}},
  \bibinfo{author}{\bibfnamefont{V.~P.} \bibnamefont{Kochereshko}},
  \bibinfo{author}{\bibfnamefont{G.~V.} \bibnamefont{Astakhov}},
  \bibinfo{author}{\bibfnamefont{D.~R.} \bibnamefont{Yakovlev}},
  \bibinfo{author}{\bibfnamefont{W.}~\bibnamefont{Ossau}},
  \bibinfo{author}{\bibfnamefont{J.}~\bibnamefont{Nurnberger}},
  \bibinfo{author}{\bibfnamefont{W.}~\bibnamefont{Faschinger}},
  \bibinfo{author}{\bibfnamefont{G.}~\bibnamefont{Landwehr}},
  \bibinfo{author}{\bibfnamefont{T.}~\bibnamefont{Wojtowicz}},
  \bibinfo{author}{\bibfnamefont{G.}~\bibnamefont{Karczewski}},
  \bibnamefont{et~al.}, \bibinfo{journal}{Physica Status Solidi B-Basic
  Research} \textbf{\bibinfo{volume}{227}}, \bibinfo{pages}{343}
  (\bibinfo{year}{2001}).

\bibitem[{\citenamefont{Ramon et~al.}(2003)\citenamefont{Ramon, Mann, and
  Cohen}}]{ramon:2003}
\bibinfo{author}{\bibfnamefont{G.}~\bibnamefont{Ramon}},
  \bibinfo{author}{\bibfnamefont{A.}~\bibnamefont{Mann}}, \bibnamefont{and}
  \bibinfo{author}{\bibfnamefont{E.}~\bibnamefont{Cohen}},
  \bibinfo{journal}{Physical Review B} \textbf{\bibinfo{volume}{67}},
  \bibinfo{pages}{45323} (\bibinfo{year}{2003}).

\bibitem[{\citenamefont{Combescot et~al.}(2005)\citenamefont{Combescot,
  Tribollet, Karczewski, Bernardot, Testelin, and Chamarro}}]{combescot:2005}
\bibinfo{author}{\bibfnamefont{M.}~\bibnamefont{Combescot}},
  \bibinfo{author}{\bibfnamefont{J.}~\bibnamefont{Tribollet}},
  \bibinfo{author}{\bibfnamefont{G.}~\bibnamefont{Karczewski}},
  \bibinfo{author}{\bibfnamefont{F.}~\bibnamefont{Bernardot}},
  \bibinfo{author}{\bibfnamefont{C.}~\bibnamefont{Testelin}}, \bibnamefont{and}
  \bibinfo{author}{\bibfnamefont{M.}~\bibnamefont{Chamarro}},
  \bibinfo{journal}{Europhysics Letters} \textbf{\bibinfo{volume}{71}},
  \bibinfo{pages}{431} (\bibinfo{year}{2005}).

\bibitem[{\citenamefont{Wojtowicz et~al.}(1998)\citenamefont{Wojtowicz,
  Kutrowski, Karczewski, and Kossut}}]{wojtowicz:1998}
\bibinfo{author}{\bibfnamefont{T.}~\bibnamefont{Wojtowicz}},
  \bibinfo{author}{\bibfnamefont{M.}~\bibnamefont{Kutrowski}},
  \bibinfo{author}{\bibfnamefont{G.}~\bibnamefont{Karczewski}},
  \bibnamefont{and} \bibinfo{author}{\bibfnamefont{J.}~\bibnamefont{Kossut}},
  \bibinfo{journal}{Applied Physics Letters} \textbf{\bibinfo{volume}{73}},
  \bibinfo{pages}{1379} (\bibinfo{year}{1998}).

\bibitem[{\citenamefont{Sergeev and Suris}(2001)}]{sergeev:2001}
\bibinfo{author}{\bibfnamefont{R.~A.} \bibnamefont{Sergeev}} \bibnamefont{and}
  \bibinfo{author}{\bibfnamefont{R.~A.} \bibnamefont{Suris}},
  \bibinfo{journal}{Physics of the Solid State} \textbf{\bibinfo{volume}{43}},
  \bibinfo{pages}{746} (\bibinfo{year}{2001}).

\bibitem[{\citenamefont{Andreani et~al.}(1998)\citenamefont{Andreani,
  Panzarini, Kavokin, and Vladimirova}}]{andreani:1998}
\bibinfo{author}{\bibfnamefont{L.~C.} \bibnamefont{Andreani}},
  \bibinfo{author}{\bibfnamefont{G.}~\bibnamefont{Panzarini}},
  \bibinfo{author}{\bibfnamefont{A.~V.} \bibnamefont{Kavokin}},
  \bibnamefont{and} \bibinfo{author}{\bibfnamefont{M.~R.}
  \bibnamefont{Vladimirova}}, \bibinfo{journal}{Phys. Rev. B}
  \textbf{\bibinfo{volume}{57}}, \bibinfo{pages}{4670} (\bibinfo{year}{1998}).

\bibitem[{\citenamefont{Schreier}(1992)}]{schreier:1992}
\bibinfo{author}{\bibfnamefont{F.}~\bibnamefont{Schreier}},
  \bibinfo{journal}{Journal of Quantitative Spectroscopy \& Radiative Transfer}
  \textbf{\bibinfo{volume}{48}}, \bibinfo{pages}{743} (\bibinfo{year}{1992}).

\bibitem[{\citenamefont{Choi et~al.}(1997)\citenamefont{Choi, Kim, Yoo, Aspnes,
  Miotkowski, and Ramdas}}]{choi:1997}
\bibinfo{author}{\bibfnamefont{S.~G.} \bibnamefont{Choi}},
  \bibinfo{author}{\bibfnamefont{Y.~D.} \bibnamefont{Kim}},
  \bibinfo{author}{\bibfnamefont{S.~D.} \bibnamefont{Yoo}},
  \bibinfo{author}{\bibfnamefont{D.~E.} \bibnamefont{Aspnes}},
  \bibinfo{author}{\bibfnamefont{I.}~\bibnamefont{Miotkowski}},
  \bibnamefont{and} \bibinfo{author}{\bibfnamefont{A.~K.}
  \bibnamefont{Ramdas}}, \bibinfo{journal}{Applied Physics Letters}
  \textbf{\bibinfo{volume}{71}}, \bibinfo{pages}{249} (\bibinfo{year}{1997}).

\bibitem[{\citenamefont{Andre and Dang}(1997)}]{andre:1997}
\bibinfo{author}{\bibfnamefont{R.}~\bibnamefont{Andre}} \bibnamefont{and}
  \bibinfo{author}{\bibfnamefont{L.~S.} \bibnamefont{Dang}},
  \bibinfo{journal}{Journal of Applied Physics} \textbf{\bibinfo{volume}{82}},
  \bibinfo{pages}{5086} (\bibinfo{year}{1997}).

\bibitem[{\citenamefont{Benhlal et~al.}(1999)\citenamefont{Benhlal, Strauch,
  Granger, and Triboulet}}]{benhlal:1999}
\bibinfo{author}{\bibfnamefont{J.~T.} \bibnamefont{Benhlal}},
  \bibinfo{author}{\bibfnamefont{K.}~\bibnamefont{Strauch}},
  \bibinfo{author}{\bibfnamefont{R.}~\bibnamefont{Granger}}, \bibnamefont{and}
  \bibinfo{author}{\bibfnamefont{R.}~\bibnamefont{Triboulet}},
  \bibinfo{journal}{Optical Materials} \textbf{\bibinfo{volume}{12}},
  \bibinfo{pages}{143} (\bibinfo{year}{1999}).

\bibitem[{\citenamefont{Hlidek et~al.}(2001)\citenamefont{Hlidek, Bok, Franc,
  and Grill}}]{hlidek:2001}
\bibinfo{author}{\bibfnamefont{P.}~\bibnamefont{Hlidek}},
  \bibinfo{author}{\bibfnamefont{J.}~\bibnamefont{Bok}},
  \bibinfo{author}{\bibfnamefont{J.}~\bibnamefont{Franc}}, \bibnamefont{and}
  \bibinfo{author}{\bibfnamefont{R.}~\bibnamefont{Grill}},
  \bibinfo{journal}{Journal of Applied Physics} \textbf{\bibinfo{volume}{90}},
  \bibinfo{pages}{1672} (\bibinfo{year}{2001}).

\bibitem[{\citenamefont{Esser et~al.}(2002)\citenamefont{Esser, Yayon, and
  Bar-Joseph}}]{esser:2002}
\bibinfo{author}{\bibfnamefont{A.}~\bibnamefont{Esser}},
  \bibinfo{author}{\bibfnamefont{Y.}~\bibnamefont{Yayon}}, \bibnamefont{and}
  \bibinfo{author}{\bibfnamefont{I.}~\bibnamefont{Bar-Joseph}},
  \bibinfo{journal}{Physica Status Solidi B} \textbf{\bibinfo{volume}{234}},
  \bibinfo{pages}{266} (\bibinfo{year}{2002}).

\bibitem[{\citenamefont{Dupertuis}(2006)}]{dupertuis:2006}
\bibinfo{author}{\bibfnamefont{M.-A.} \bibnamefont{Dupertuis}}
  (\bibinfo{year}{2006}), \bibinfo{note}{private communication}.

\bibitem[{\citenamefont{Berney et~al.}(2007)\citenamefont{Berney,
  Portella-Oberli, and Deveaud}}]{berney:2007}
\bibinfo{author}{\bibfnamefont{J.}~\bibnamefont{Berney}},
  \bibinfo{author}{\bibfnamefont{M.~T.} \bibnamefont{Portella-Oberli}},
  \bibnamefont{and} \bibinfo{author}{\bibfnamefont{B.}~\bibnamefont{Deveaud}},
  \bibinfo{journal}{Cond. Matt} p. \bibinfo{pages}{\#arXiv:07082330}
  (\bibinfo{year}{2007}).

\bibitem[{\citenamefont{Savona}(1999)}]{cargese:1999}
\bibinfo{author}{\bibfnamefont{V.}~\bibnamefont{Savona}}, in
  \emph{\bibinfo{booktitle}{Confined photon systems: Fundamentals and
  applications}} (\bibinfo{publisher}{Springer Verlag, Berlin, New York},
  \bibinfo{year}{1999}), pp. \bibinfo{pages}{173--242}.

\bibitem[{\citenamefont{Born and Wolf}(1999)}]{born:1999}
\bibinfo{author}{\bibfnamefont{M.}~\bibnamefont{Born}} \bibnamefont{and}
  \bibinfo{author}{\bibfnamefont{E.}~\bibnamefont{Wolf}},
  \emph{\bibinfo{title}{Principles of Optics}} (\bibinfo{publisher}{Cambridge
  University Press}, \bibinfo{year}{1999}), \bibinfo{edition}{7th} ed.

\bibitem[{\citenamefont{Esser et~al.}(2000)\citenamefont{Esser, Runge,
  Zimmermann, and Langbein}}]{esser:2000}
\bibinfo{author}{\bibfnamefont{A.}~\bibnamefont{Esser}},
  \bibinfo{author}{\bibfnamefont{E.}~\bibnamefont{Runge}},
  \bibinfo{author}{\bibfnamefont{R.}~\bibnamefont{Zimmermann}},
  \bibnamefont{and} \bibinfo{author}{\bibfnamefont{W.}~\bibnamefont{Langbein}},
  \bibinfo{journal}{Physical Review B} \textbf{\bibinfo{volume}{62}},
  \bibinfo{pages}{8232} (\bibinfo{year}{2000}).

\bibitem[{\citenamefont{Chandrasekhar}(1944)}]{chandrasekhar:1944}
\bibinfo{author}{\bibfnamefont{S.}~\bibnamefont{Chandrasekhar}},
  \bibinfo{journal}{Astrophysical Journal} \textbf{\bibinfo{volume}{100}},
  \bibinfo{pages}{176} (\bibinfo{year}{1944}).

\end{thebibliography}
\end{document}